\documentclass[aps,prb,manuscript]{revtex4}
\usepackage{epsfig}
\usepackage{graphics}
\usepackage{bm}
\begin{document}

\title{Dynamic and static control of the optical phase of
guided $p$-polarized light for near-field focusing at large
angles of incidence}

\author{Danhong Huang$^{1}$, M. Michelle Easter$^{2}$, L. David Wellems$^{1}$, Henry Mozer$^{1}$, Godfrey Gumbs$^{2}$, D. A. Cardimona$^{1}$, and A. A. Maradudin$^{3}$}

\affiliation{$^{1}$Air Force Research Laboratory, Space Vehicles
Directorate, Kirtland Air Force Base, NM 87117, USA\\
$^{2}$Department of Physics and Astronomy, Hunter College at the City University of New York, 695 Park Avenue New York, NY 10065, USA\\
$^{3}$Department of Physics and Astronomy and Institute for Surface and Interface Science,
University of California, Irvine, CA 92697, USA}

\date{\today}

\begin{abstract}
Both dynamic and static approaches are proposed and investigated for controlling the optical phase of a $p$-polarized light wave that is guided through a surface-patterned metallic structure with subwavelength features. For dynamic control, the important role of photo-excited electrons in a slit-embedded atomic system with field-induced transparency (FIT) is discovered within a narrow frequency window for modulating the intensity of focused transmitted light in the near-field region. Based on the electromagnetic coupling. This is facilitated by surface plasmons between the two FIT-atom embedded slits. The near-field distribution may be adjusted by employing a symmetric (or asymmetric) slit configuration as well as  by either a small or large  slit separation. Furthermore, the cross-transmission of a light beam is predicted as a result of strong coupling between optical transitions within embedded FIT atoms and surface plasmons. For static control, the role of surface curvature is obtained for focused transmitted light passing through a Gaussian-shaped metallic microlens embedded with a linear array of slits, in addition to a negative light-refraction pattern, which is associated with higher-diffraction modes of light, under a large angle of incidence in the near-field region. Most interesting, however, this anomalous negative light-refraction pattern becomes suppressible with leaked higher-order waveguide modes of light passing through a very thin film. At the same time, it is also suppressible with a reinforced reflection at the left foothill of a Gaussian-shaped slit array for the forward-propagating surface-plasmon wave at large angles of incidence. A prediction is given for near-field focusing of light with its sharpness dynamically controlled by the frequency of the light in a very narrow window. Moreover, a different scheme based on Green's second integral identity is proposed for overcoming a difficulty in calculating the near-field distribution very close to a surface by means of a finite-difference-time-domain method.
\end{abstract}


\maketitle

\section{Introduction}

Dipole-coupling based optical transition of electrons in an atom is usually very sharp and strong under a resonant frequency of the light. As a result, the coherent electron dynamics occurs only within an extremely-narrow frequency window. If a laser is used to resonantly excite a multi-level atomic system, the {\em quantum interference} will simultaneously occur in the optical absorption spectrum as well as in the spontaneous emission spectrum at the same time. A quantum electrodynamic theory, that treats both photons and electrons in second quantization, is compulsory for looking into the radiative decay of excited electrons in an atomic system\,\cite{milonni}. In addition to the well-known diagonal radiative-decay time of atoms, there exists another less-known {\em off-diagonal radiative-decay coupling} (ODRDC) that becomes significant as two or more electronic transition energies are very close.\,\cite{ref5,ref3} Explicitly, the ODRDC phenomenon is associated with a nearly-resonant reabsorption of a spontaneously emitted photon from one radiative decay process by another upward stimulated optical transition of electrons. Consequently, any two nearly-degenerate electronic transitions in the presence of ODRDC will be coupled and interfere with each other, leading to a zero absorption line\,\cite{ref5,ref3,cardimona1,cardimona2,ref4} in the spectral region where two absorption peaks are overlapped.
\medskip

The collective excitation of electrons in metals can couple to light to form surface-plasmon polaritons through dielectric-function dependent boundary conditions for electromagnetic fields on both sides of an interface. The incoherent dynamics of excited surface-plasmon polaritons usually shows up in a broad frequency range owing to the high optical loss of metals. In recent years, surface-plasmon polaritons\,\cite{raether} and local surface plasmons\,\cite{bookg} as well, which are localized at a dielectric-metal interface, have acquired a great deal of attention\,\cite{vidal}. The experimentally observed extraordinarily-high transmission of a $p$-polarized light beam passing through a two-dimensional hole array embedded on a metal film in the subwavelength regime\,\cite{ebbesen,ebbesen1,ebbesen2} has been proved to depend on both the array period and film thickness in a sensitive way\,\cite{lin,lin2}. In addition, studies on the surface-plasmon mediated light transmission by a subwavelength structure on a metal surface have also attracted a great deal of attention\,\cite{bookaa,maradudin}. If a single slit is perforated on an optically-opaque metal film deposited on a dielectric substrate, it is discovered that the induced surface plasmons on the front side of the film will be coupled to the backside by intra-slit interferences\,\cite{vidal,Perez}. It is also found that not only the intra-slit interference but also the inter-slit interference in a slit array is a vital component for the transmission of the excited surface plasmons passing through the slits\,\cite{ref1}. Furthermore, the inter-slit interference is shown to relate only to the surface-wave nature and not necessarily to the surface-plasmon wave, because this interference survives even for a perfect electric conductor.
\medskip

It is known that surface patterning of a metal film at the nanoscale, {\em e.g.\/}, slits, holes, and even surface corrugations, can provide flexible control of light incident on it, for example, flat metallic lenses have been constructed with nanoslits perforated on a thin metal film to demonstrate beam deflection and focusing based on specific phase retardation controlled by modulated slit depths\,\cite{ref21}, slit widths\,\cite{ref22}, slit separations\,\cite{ref23}, and slit dielectric filling\,\cite{ref1}. Moreover, surface plasmons have been identified as the essential ingredient in these phenomena, and the Fabry-Perot resonance of surface plasmons contributes greatly to the enhanced transmission and localized electric-field in the slit perforated metallic film\,\cite{ref24,ref25}. The configurations of a single aperture smaller than the incident wavelength and an array of nanoslits perforated on a metallic film have attracted a great amount of attention due not only to extraordinary optical transmission of light\,\cite{ref26}, but also to their beaming effect in the far field\,\cite{ref27,ref28,ref29,ref30,ref31} and the``lens-like'' properties\,\cite{ref22,ref32} for focusing normally-incident light in the intermediate region within a certain frequency range.
In the near-field region, the propagation of a light wave guided through a metallic structure in the subwavelength regime is determined by light diffraction, instead of by Snell's law for a light ray. The optical phase of a guided light wave is a product of the host-material refractive index and the geometrical distance traveled by the light inside the host. The frequency-dependent refractive-index function reflects the dynamics of photo-excited electrons in the time domain, which can be {\em dynamically} controlled through an optical resonance between a pair of discrete energy levels in addition to a weak Kerr optical nonlinearity, such as embedded atoms or quantum dots, within a very narrow frequency window. For this dynamic-control approach, strong nonlinear coupling of electrons to an incident light needs to be taken into account, {\em e.g.\/}, the coupled Maxwell-Bloch equations for including nonlinear optical response of photo-excited electrons inside atoms to a spatially-inhomogeneous total electromagnetic field have to be solved. On the other hand, the geometrical distance traveled by the light can be {\em statically} controlled by changing the surface profile of a host material at its interface with either air or a substrate, {\em e.g.\/}, a Gaussian-shaped metallic microlens embedded with a finite nanoslit array. For this static-control approach, only the Maxwell equations with the inclusion of a linear optical response of photo-excited electrons in conductive materials are required.
\medskip

The {\em motivation} for the current research is stated as follows. First, it is well know that the slow and incoherent dynamics of surface-plasmon excitation can produce a very strong near field close to a surface and a broad transmission peak in the spectrum. On the other hand, the fast and coherent dynamics originated from stimulated transitions of electrons inside atoms can lead to a strong absorption peak in which a very narrow transparency window is buried. Therefore, the combination of these two quite different dynamics is expected to introduce a new mechanism for near-field focusing of light with its sharpness {\em dynamically} controlled by the frequency of the light in an extremely narrow window in the spectrum. Second, for a plane-wave guided through a metallic nanoslit array in the subwavelength regime, it is also well known that only the near field can survive as the incident angle of light becomes very large. Although a flat lens composed of nanoslits perforated on a thin metal film can be applied to focusing normally-incident light, a long tail along the lens axis in the light-focusing pattern is unavoidable. The use of a curved nanoslit array embedded on a metal film, on the other hand, can provide a very sharp and configuration-dependent focus spot without a longitudinal tail even at large angles of incidence. In this sense, the near-field focusing of a plane-wave light can be controlled {\em statically} mainly through a change of the surface curvature of a metallic lens. As far as the near-field distribution very close to a surface is concerned, the commonly-employed simple finite-difference-time-domain method drops memory effect in the time domain and tends to leave out details in the field-distribution pattern very close to the surface, which requires inclusion of very large transverse wave numbers (or very fine spatial meshes in the direction parallel to the surface) in numerical calculations. However, the analytic calculations based on Green's second integral identity in this work do not suffer from this drawback.
\medskip

In this paper, we investigate both dynamic and static control of the optical phase for a $p$-polarized light wave guided through a subwavelength surface-patterned metallic structure. For  dynamic control, we explore the coherent dynamics of photo-excited electrons in a slit-embedded atomic system with field-induced transparency (FIT) within a narrow frequency window, which is further superposed on the incoherent dynamics of surface plasmons in a broad frequency range. For the static control, we investigate the effect of surface curvature on focused transmitted light passing through a Gaussian-shaped metallic microlens embedded with a finite array of nanoslits under a very large angle of incidence in the near-field region.
\medskip

The rest of the paper is organized as follows. In Sec.\,\ref{sec2}, we introduce two model systems, respectively, for dynamic and static control of the optical phase of a guided $p$-polarized light wave. In Sec.\,\ref{sec3}, numerical examples are presented and discussed in detail for dynamic and static optical-phase control in these two model systems. Finally, a brief summary of the main results of this paper is given in Sec.\,\ref{sec4}.

\section{Model Systems for Optical-Phase Control}
\label{sec2}

In this section, we will introduce two model systems associated with the dynamic and static control of the optical phases, respectively. Based on these two introduced model systems, numerical calculations
are performed to demonstrate both the coherent electron dynamics in embedded atoms randomly-distributed within a nanoslit and the incoherent dynamics associated with interference effects of resonantly excited
surface-plasmon waves. Also shown is the suppression of interference by reinforced reflection at the ``foothill'' of a curved slit array under large angles of incidence.

\subsection{Dynamic approach}
\label{sec2-1}

\begin{figure}[p]
\centering
\includegraphics[width=0.7\textwidth]{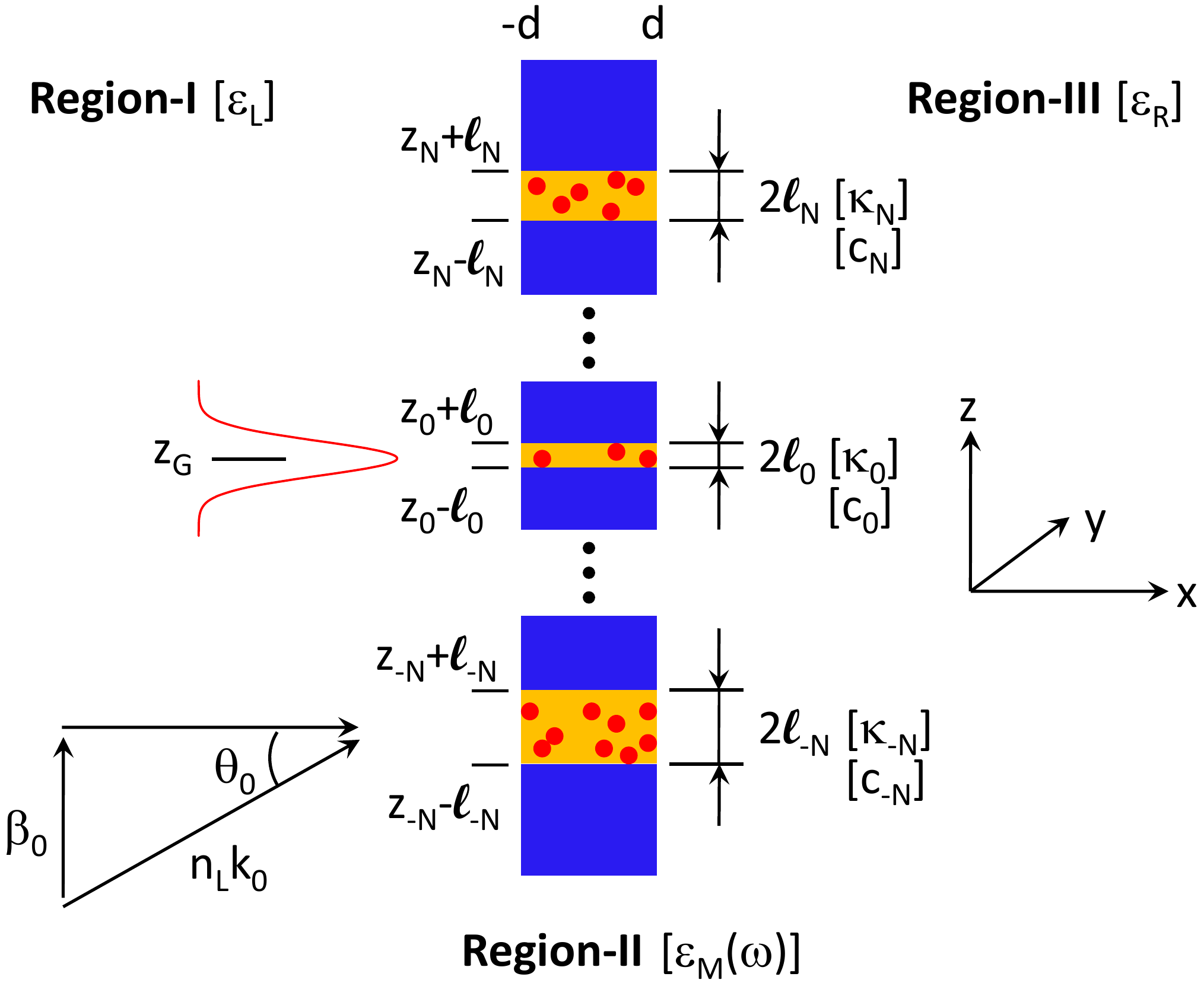}
\caption{\label{f1}
(Color online) Schematic illustration for a $z$ direction slit array
(blue) which extends in the $y$ direction. In our notation, $z_j$ and $2\ell_j$
denote the center position and the slit width of the $j$-th slit, respectively,
with $j=0,\,\pm 1,\,\cdots,\,\pm N$. The regions on the left- and
right-hand sides of the slits are indicated as Region I and Region III,
respectively, with real dielectric constants $\epsilon_{\rm L}$ and
$\epsilon_{\rm R}$. The region for the slit array is denoted as Region II,
and slits are filled with medium (orange) having  dielectric constant
$\kappa_j$ (real) for $j=0,\,\pm 1,\,\cdots,\,\pm N$.
Randomly-distributed atoms (red dots) are embedded inside each slit-filled dielectric medium with a concentration $c_j$
for $j=0,\,\pm 1,\,\cdots,\,\pm N$ and
described by an effective, frequency-dependent and complex dielectric function. The
depth of slits in the $x$ direction is $2d$, and
$\epsilon_{\rm M}(\omega)$ represents the dielectric function of the metal
film containing slits. A Gaussian beam is incident on the slit array
from the left-hand side with the angle $\theta_0$ of incidence and at a center
position $z=z_{\rm G}$. The incident wave number is
$\sqrt{\epsilon_{\rm L}}\,k_0$ and $\beta_0$ is the incident wave vector
along the $z$ direction.
}
\end{figure}

By taking into account of the dynamics of photo-excited electrons within embedded atoms, the model system introduced for dynamically controlling the optical phase is illustrated in Fig.\,\ref{f1}.
The complete theory beyond a so-called ``diagonal'' approximation\,\cite{vidal,Perez,vidal2}, which is based on the modal expansion and the surface-impedance-boundary condition,
for calculating the spatial distribution of the total electromagnetic field as well as the transmission and reflection spectra in such a model system has already been reported in Ref.\,[\onlinecite{ref1}].
Here, the dielectric constant $\kappa_j$ for the nanoslit-filled material in Ref.\,[\onlinecite{ref1}] should be replaced by a frequency-dependent effective one with the inclusion of
randomly-distributed atoms (random in position but with the same locally-spherical symmetry)
embedded within the nanoslit-filled dielectric medium. Explicitly, we require the substitution $\kappa_j\to\epsilon_j(\omega)\equiv\kappa_j+\alpha^j_{\rm L}(\omega)$ in Ref.\,[\onlinecite{ref1}] with the
Lorentz ratio of photo-excited electrons given by\,\cite{ref3}

\begin{equation}
\alpha^j_{\rm L}(\omega)=\frac{c_j\hbar}{2\epsilon_0[E^j_{\rm eff}]^2}\,\left[\Omega^j_{13}\,{\rho}^j_{31}(\omega)+\Omega^j_{12}\,{\rho}^j_{21}(\omega)\right]\ ,
\label{e1}
\end{equation}
where $c_j$ is the concentration of embedded atoms in the $j$th nanoslit, $\hbar\omega$ is the incident photon energy,
$\Omega^j_{\nu\mu}=2d_{\nu\mu}E^j_{\Theta}/\hbar$ for $\nu\neq\mu=1,\,2,\,3$ (three-level system)
is the Rabi frequency, ${\bf d}_{\nu\mu}$ is the dipole moment for the optical transition
of electrons from a lower energy level $|\nu>$ to a higher energy level $|\mu>$,

\begin{equation}
E^j_{\Theta}=\frac{1}{{\cal S}_j}\int_{{\cal S}_j} d^2{\bf r}_\|\,E({\bf r}_\|)\,\Theta({\bf r}_\|)\ ,\ \ \ \ \mbox{and}\ \ \ \ E^j_{\rm eff}=\frac{1}{{\cal S}_j}\int_{{\cal S}_j} d^2{\bf r}_\|\,E({\bf r}_\|)\ .
\label{e2}
\end{equation}
In Eq.\,(\ref{e2}), the spatial averages of the product $E({\bf r}_\|)\,\Theta({\bf r}_\|)$ and $E({\bf r}_\|)$ itself are performed within the $j$th-nanoslit with an area ${\cal S}_j$,
${\bf r}_\|$ is a two-dimensional position vector within the $xz$-plane,  $E({\bf r}_\|)$ is the magnitude of the electric-field component,
and $\Theta({\bf r}_\|)$ represents its structure factor in the $xz$-plane.
\medskip

In order to relate the electric-field vector ${\bf E}({\bf r}_\|)$ to the solution of the Maxwell equations in Ref.\,[\onlinecite{ref1}], we consider a $p$ polarization for the incident light
as an example for this paper. Therefore, the total magnetic-field vector ${\bf H}({\bf r}_\|)$ in the direction perpendicular to the $xz$-plane,
as shown in Fig.\,\ref{f1}, can be simply written as ${\bf H}({\bf r}_\|)=[0,\,H_y(x,z),\,0]$. This gives rise to an expression for the electric-field vector

\begin{equation}
{\bf E}({\bf r}_\|)=\frac{i}{\omega\epsilon_0\epsilon_j(\omega)}\,\left[-\frac{\partial H_y(x,z)}{\partial z},\,0,\,\frac{\partial H_y(x,z)}{\partial x}\right]\ ,
\label{e3}
\end{equation}
as well as its structure factor, given by,

\[
\Theta({\bf r}_\|)=\frac{1}{\sqrt{3}}\,\frac{\partial H_y(x,z)/\partial x-\partial H_y(x,z)/\partial z}{\sqrt{[\partial H_y(x,z)/\partial x]^2+[\partial H_y(x,z)/\partial z]^2}}\ .
\]
\medskip

Finally, the density-matrix elements ${\rho}^j_{\mu\nu}(\omega)$ for the $j$th-nanoslit, which are introduced in Eq\,(\ref{e1}) for $\mu,\,\nu=1,\,2,\,3$,
can be calculated by assuming a three-level atomic system with the property of
field-induced transparency\,\cite{ref5,ref4}, as shown in Fig.\,\ref{f2}(a).
The details, which are based on the optical Bloch equations with quantum interference from the ODRDC,
for calculating the density-matrix elements ${\rho}^j_{\mu\nu}(\omega)$ in such a system have already been formulated in Refs.\,[\onlinecite{ref3},\,\onlinecite{ref4}].
The iterations are required for solving the coupled nonlinear Maxwell-Bloch equations
and getting the self-consistent solutions with respect to $H_y(y,z)$ and ${\rho}^j_{\mu\nu}(\omega)$ at the same time.

\subsection{Static approach}
\label{sec2-2}

By including a nontrivial surface profile to the interface of a host material, the model system introduced for statically controlling the optical phase is illustrated in Fig.\,\ref{f6}.
The in-detail formalism, which is based on Green's second integral identity in the $xz$-plane\,\cite{ref6},
for calculating the spatial distribution of the total electromagnetic field as well as the transmission and reflection spectra in such a model system has already been reported in Refs.\,[\onlinecite{ref6,ref2}].
The top surface profile in Ref.\,[\onlinecite{ref2}] should be replaced by

\begin{equation}
\xi_1(x)=\left\{\begin{array}{cc}
{\cal A}_0\,\exp\left(-\frac{x^2}{\sigma^2_0}\right)\,\sum\limits_{j=-N}^N\,\exp\left[-\frac{(x-jd)^2}{b^2}\right]\ , &\ \ \ \ \ \ \ \ \mbox{for $|x|\leq D_0/2$}\\
Q_0\ , &\ \ \ \ \ \ \ \ \mbox{for $|x|>D_0/2$}
\end{array}\right.\ ,
\label{e4}
\end{equation}
where $j$ is the nanoslit index, $d$ is the array period, $2b$ represents the width of nanoslit-separation layer, $2N+1$ is the total number of nanoslits with width $\sim(d-2b)$. In addition, $D_0$ is the microlens aperture size, $Q_0+a\approx Q_0$ is the metal film thickness, ($a$ is much smaller than the incident wavelength $\lambda_0$ for strong photon tunneling),
${\cal A}_0$ and $2\sigma_0$ are the amplitude and width for the nanoslit-depth modulation.
Additionally, the bottom surface profile in Ref.\,[\onlinecite{ref2}] is replaced by a flat one, {\em i.e.\/}, $\xi_2(x)=-a$.
For $|x|\ll 1$, the quadratic term in the expansion of $\xi_1(x)$ in Eq.\,(\ref{e4}) leads to an effective curvature-based focal distance $f_0\sim\sigma_0^2/(2{\cal A}_0)$ measured from the exit flat surface.

\section{Numerical Results and Discussions}
\label{sec3}

In our numerical calculations, the metal is chosen to be silver (gold) for dynamic (static) control,
and its dielectric functions $\epsilon_{\rm M}(\omega)$ with different frequency $\omega$ of the light are taken from the work by Johnson and Christy\,\cite{table}.

\subsection{Quantum-interference effect in dynamic control}
\label{sec3-1}

\begin{figure}[p]
\centering
\includegraphics[width=0.7\textwidth]{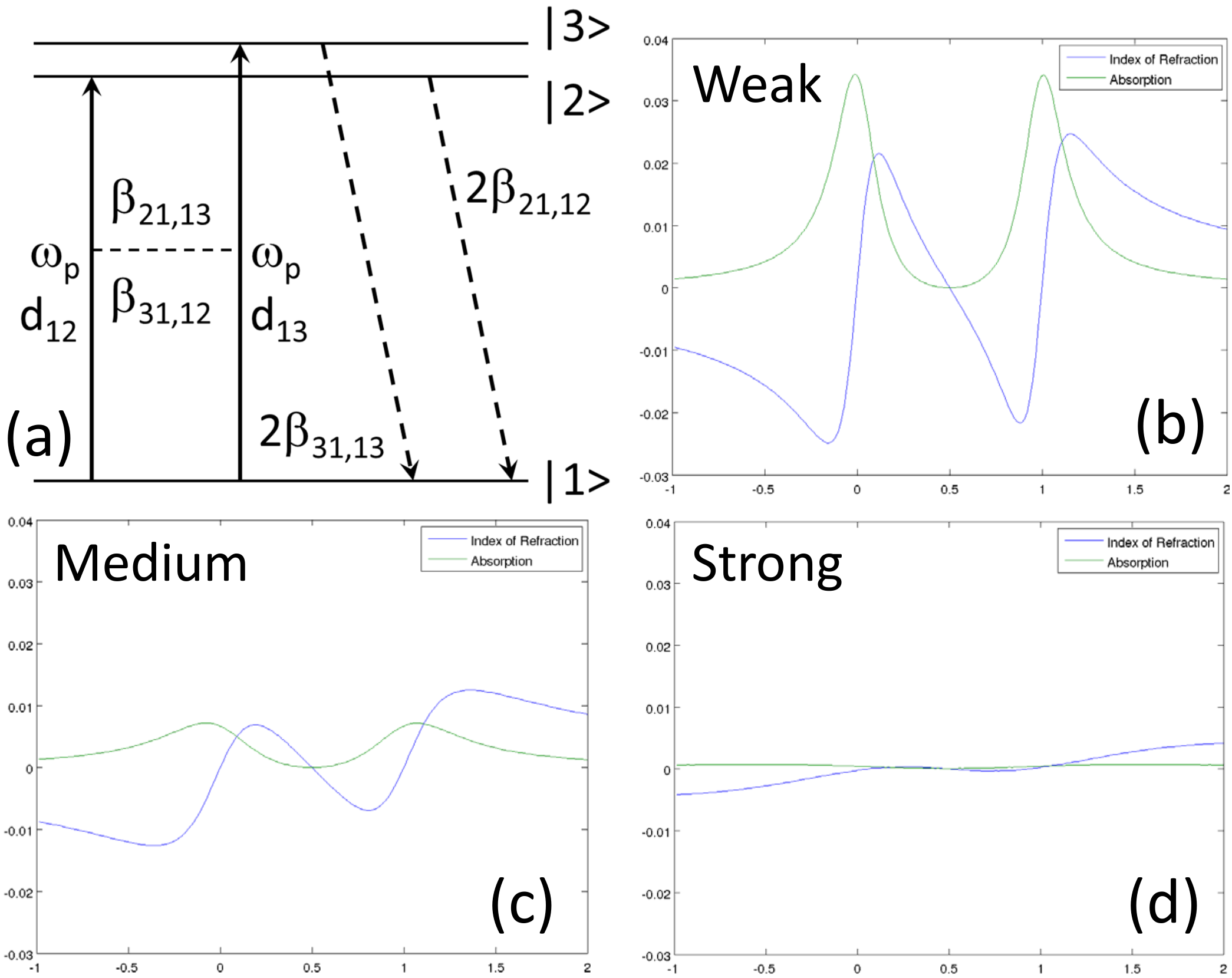}
\caption{\label{f2}
(Color online) Schematic illustration for the effect due to
off-diagonal radiative-decay coupling in the bare-atom picture
[in (a)] and the scaled
$\alpha_{\rm L}(\omega_{\rm p})\,(2\epsilon_0\omega_{32}E_{\rm p}^2/c_{\rm at}\hbar\Omega_{\rm p}^2)$,
defined through Eq.\,(\ref{e1}), as functions of the detuning $\delta=(\omega_{\rm p}-\omega_{21})/\omega_{32}$ in the other panels (b), (c) and (d), where $c_{\rm at}$ is the atom concentration,
$\hbar\omega_{ji}=\varepsilon_j-\varepsilon_i$ with $\omega_{32}\ll\omega_{21}$ and $\varepsilon_j$ for $j=1,\,2,\,3$ stand for the energies of three labeled states in (a).
The notations $|1>$, $|2>$ and $|3>$ in (a) represent different energy states in the bare-atom
picture, $\omega_{\rm p}$ is the probe-field frequency and $d_{12},\,d_{13}$ are nonzero optical dipole moments,
and $\beta_{ij,mn}$ denotes the real part of both the diagonal ($\beta_{21,12}$ and $\beta_{31,13}$) and off-diagonal ($\beta_{21,13}$ and $\beta_{31,12}$) radiative-decay rates.
In (b), (c) and (d) both the real (index of refraction) and the imaginary (absorption ) parts of $\alpha_{\rm L}(\omega_{\rm p})$ have been presented for the pure atomic system in (a)
with three different given values for the probe-field amplitude $E_{\rm p}$, {\em i.e.\/}, $\Omega_{\rm p}/\omega_{32}=0.067$ (weak), $0.2$ (medium) and $0.67$ (strong), where $\Omega_{\rm p}=d_{13}\,E_{\rm p}/\hbar$,
where ${\bf d}_{13}={\bf d}_{12}$ and ${\bf d}_{23}=0$ are assumed.
}
\end{figure}

\begin{figure}[p]
\centering
\includegraphics[width=0.7\textwidth]{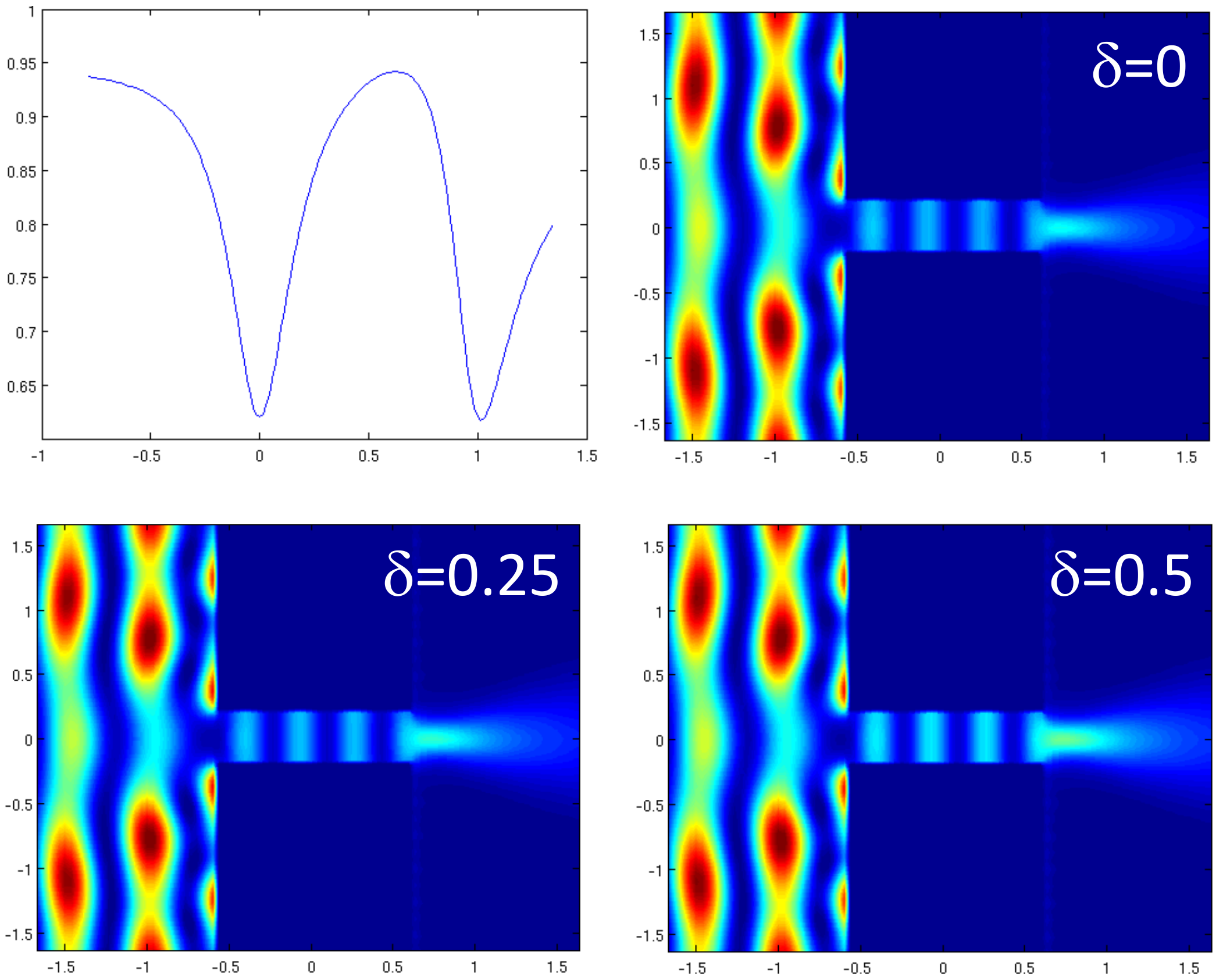}
\caption{\label{f3}
(Color online)  Transmission coefficient calculated self-consistently
as a function of $\delta$ in the upper-left panel. Also, shown are
color maps for $|H_y(x,z)|^2$ with a single slit. The slit is
 filled with a dielectric  material ($\kappa_0=2.25$) which is
 embedded with randomly-distributed atoms in the rest of the panels
 with $\delta=0$ (upper-right), $0.25$ (lower-left) and
 $0.5$ (lower-right), respectively. The width and position of
 the slit are $\ell_0=0.2\,\mu$m and $z_0=0$. Here, the
 incident-light wavelength is $\lambda_0=990$\,nm and the
 $p$-polarized plane-wave amplitude is $\mu_0H_0=0.5$\,G.
}
\end{figure}

\begin{figure}[p]
\centering
\includegraphics[width=0.9\textwidth]{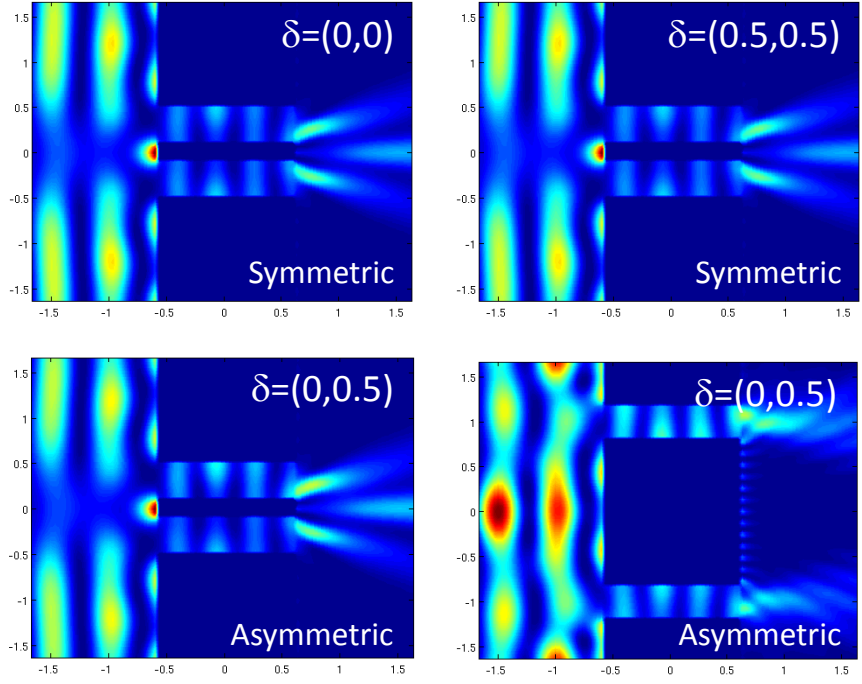}
\caption{\label{f4}
(Color online) Self-consistently determined double-slit color
maps for $|H_y(x,z)|^2$ filled with dielectric material ($\kappa_{-1}=\kappa_1=2.25$) which are both embedded with
randomly-distributed atoms. The upper panels correspond to a
symmetry configuration having the same embedded atoms in both
slits, {\em i.e.\/}, $\delta_{-1}=\delta_1=0$ (upper left) and $\delta_{-1}=\delta_1=0.5$ (upper-right). The lower panels are
for an asymmetric configuration with different types of embedded
atoms in two slits, {\em i.e.\/}, $\delta_{-1}=0$ and $\delta_1=0.5$
for the lower and upper slits, respectively. In addition, for the two
lower panels, we display the results for both coupled (lower-left with
a small slit separation) and uncoupled (lower-right with a large slit
separation) slits. The widths of two slits are $\ell_{-1}=\ell_1=0.2\,\mu$m.
For the upper two and lower-left panels, the positions for these
slits are $z_{-1}=-0.25\,\mu$m and $z_1=0.25\,\mu$m, while
$z_{-1}=-1\,\mu$m and $z_1=1\,\mu$m for the lower-right panel.
Here, the incident-light wavelength is $\lambda_0=990$\,nm and
the $p$-polarized plane-wave amplitude is $\mu_0H_0=0.5$\,G.
}
\end{figure}

\begin{figure}[p]
\centering
\includegraphics[width=0.7\textwidth]{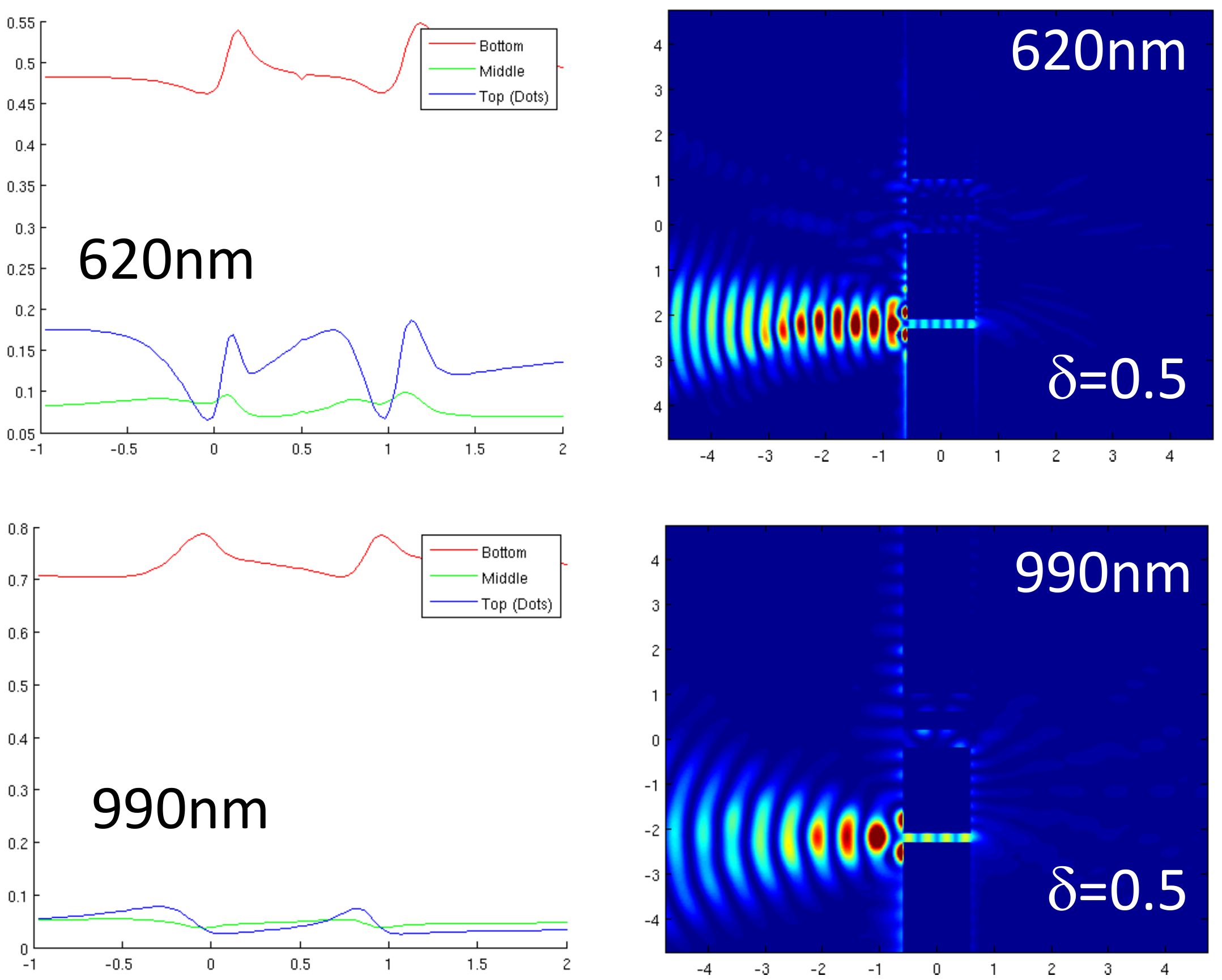}
\caption{\label{f5}
(Color online) Self-consistent calculations of direct (bottom) and cross (middle and top) triple-slit transmissions as functions of $\delta$ in the upper-left ($\lambda_0=620$\,nm) and lower-left ($\lambda_0=990$\,nm) panels.
Also plotted are their corresponding color maps for $|H_y(x,z)|^2$ with triple slits. All  slits are filled with the same dielectric material
($\kappa_{-1}=\kappa_0=\kappa_1=2.25$) but only the top slit is embedded with randomly-distributed atoms having $\delta=0.5$ for two right panels.
The widths for these three slits are $\ell_{-1}=0.1\,\mu$m and $\ell_0=\ell_1=0.2\,\mu$m, while the positions for these slits are $z_{-1}=-2.2\,\mu$m,
$z_0=0$, $z_1=0.8\,\mu$m and $z_{\rm G}=z_1$.
In addition, only the lowest slit is illuminated with light of a Gaussian beam, the incident-light wavelength is $\lambda_0=620$\,nm (upper panels) or $\lambda_0=990$\,nm (lower panels),
and the peak amplitude of a $p$-polarized Gaussian-beam is $\mu_0H_0=0.5$\,G.
}
\end{figure}

For a filled metallic nanoslit array shown in Fig.\,\ref{f1},
the parameters taken for the numerical calculations in Figs.\,\ref{f2}$-$\ref{f5} are: $\epsilon_{\rm L}=1$ for air, $\epsilon_{\rm R}=2.13$ for glass,
$2d=1.155\,\mu$m for film thickness, and $\theta_0=0^{\rm o}$ for normal incidence.
For the stand-alone three-level atomic system, we take: $\beta_{21,12}/\omega_{32}=\beta_{31,12}/\omega_{32}=\beta_{21,13}/\omega_{32}=\beta_{31,12}/\omega_{32}=0.1$, $\gamma_0/\omega_{32}=0.1$, $d_{23}=0$, $d_{12}=d_{13}$, $\delta=(\omega_{\rm p}-\omega_{21})/\omega_{32}$, $\Omega_{13}/\omega_{32}=\Omega_{12}/\omega_{32}=\Omega_{\rm p}/\omega_{32}$, $\Omega_{\rm p}=2d_{12}E_{\rm p}/\hbar$ with $E_{\rm p}$ being the probe-field amplitude.
Furthermore, for the embedded atoms within the nanoslit-filled dielectric material, we choose the same atom concentration $c_j=1\times 10^{18}$\,cm$^{-3}$ for all the nanoslits, $\hbar\omega_{32}=0.1$\,meV, and
$d_{12}/e=d_{13}/e=1$\,\AA. Other parameters as well as the changes
of these parameters will be given in the figure captions.
For all  plots in Figs.\,\ref{f3}$-$\ref{f5}, the color scale is
kept the same from $0$ (blue) to $1$ (red).
\medskip

Figure\ \ref{f2} presents the scaled Lorentz-ratio function $\alpha^{\rm at}_{\rm L}(\omega_{\rm p})\,(2\epsilon_0\omega_{32}E_{\rm p}^2/c_{\rm at}\hbar\Omega_{\rm p}^2)$
for a stand-alone atomic system with field-induced transparency, as shown in Fig.\,\ref{f2}(a),
as a function of the dimensionless detuning parameter $\delta=(\omega_{\rm p}-\omega_{21})/\omega_{32}$ for different strengths of a probe field $E_{\rm p}$, given by
$\Omega_{\rm p}/\omega_{32}=0.067$ [weak, in Fig.\,\ref{f2}(b)], $0.2$ [medium, in Fig.\,\ref{f2}(c)] and $0.67$ [strong, in Fig.\,\ref{f2}(d)].
When $E_{\rm p}$ is weak, we find a zero absorption from ${\rm Im}[\alpha^{\rm at}_{\rm L}(\omega_{\rm p})]$ in Fig.\,\ref{f2}(b)
at $\delta=0.5$, {\em i.e.\/}, the field-induced transparency (FIT) owing to quantum interference between direct and an indirect optical-transition phase
based on ODRDC\,\cite{ref3,ref4}.
This transparency is accompanied by two absorption peaks at both sides of $\delta=0.5$. In addition, corresponding changes in
${\rm Re}[\alpha^{\rm at}_{\rm L}(\omega_{\rm p})]$ with respect to these two absorption peaks are seen with a very large negative slope for
$d\{{\rm Re}[\alpha^{\rm at}_{\rm L}(\omega_{\rm p})]\}/d\omega_{\rm p}$ within a very narrow transparency window.
As $E_{\rm p}$ is gradually increased from Fig.\,\ref{f2}(c) to Fig.\,\ref{f2}(d), the strength of these two absorption peaks greatly decreases while
the width of them (power broadening) significantly increases at the same time. The peak reduction will continue with increasing $E_{\rm p}$ until the complete suppression of absorption peaks
is achieved at a saturation limit.
The results in this figure clearly demonstrate a rapid dynamic change of $\alpha^{\rm at}_{\rm L}(\omega_{\rm p})$ and a very strong probe-field nonlinearity
within an extremely-narrow frequency window $\Delta\omega_{\rm p}\sim\omega_{32}\gg\gamma_0$.
\medskip

When the FIT atomic system in Fig.\,\ref{f2}(a) is embedded into a single nanoslit, the probe field will be replaced by a guided light wave. At the same time, the electron dynamics within a narrow-frequency window for
an isolated atomic system is also inserted into a relatively broad frequency range for the dynamics of surface-plasmon polaritons.
Moreover, the FIT atoms will be coupled to surface-plasmon modes in a nanoslit.
For a very narrow and long nanoslit in the deep-subwavelength regime, we know for $p$-polarization incidence only the lowest waveguide mode can be transmitted through the nanoslit.
We display in Fig.\,\ref{f3} the self-consistently calculated transmission coefficient in the upper-left panel and color maps for $|H_y(x,z)|^2$ in the other three panels
with various detuning parameters $\delta$ (or equivalently, different FIT atoms)
in the rest three panels below the saturation limit.
A high light reflection by the front surface of a silver film,
as well as the induced surface-plasmon propagation along this surface, can be seen from Fig.\,\ref{f3}(b)$-$(d), but the pattern of $|H_y(x,z)|^2$ within the nanoslit remains the same except for
intensity variation.
It is important to note that the nanoslit has been prepared for one of the passing states satisfying $8d=|2m-1|\lambda_0/\sqrt{\kappa_0}$ with $m=0,\,\pm 1,\,\pm 2,\,\cdots$
resulted from dual-wave constructive interference\,\cite{ref1}.
Compared with the medium ($\delta=0.25$ or $\omega_{21}=2\pi c/\lambda_0-\omega_{32}/4$) and with the strongest
($\delta=0.5$ or $\omega_{21}=2\pi c/\lambda_0-\omega_{32}/2$) light transmissions in the lower-left and lower-right panels of the figure,
the transmitted light at $\delta=0$ (or $\omega_{21}=2\pi c/\lambda_0$) in the upper-right panel of the figure
is the weakest owing to maximized absorption at this particular $\delta$ value, as can be verified by the calculated transmission coefficient in the upper-left panel of this figure.
In this sense, the intensity of transmitted light can be self-modulated (or dynamically controlled) in a narrow-frequency window by embedding atoms within a nanoslit of a silver film,
and the modulation effect can be enhanced by increasing the concentration of embedded atoms in the nanoslit.
\medskip

Figure\ \ref{f4} displays the calculated color maps for $|H_y(x,z)|^2$ with double nanoslits in four different cases. In the upper-left and upper-right panels of this figure,
we consider a symmetric configuration of double nanoslits
with $\delta=0$ and $\delta=0.5$, respectively, for both nanoslits. Similar to the color maps in Fig.\,\ref{f3} for a single nanoslit,
we observe a small reduction in transmitted light through nanoslits for $\delta=0$ due to a relatively large absorption.
Because of the electromagnetic coupling between two nanoslits, a hot spot occurs in the reflection side at the front surface between two close-by nanoslits. Moreover, two candle-flame-like spots at the nanoslit exits
are significantly deformed in shape and bent down from the middle point of an individual nanoslit towards the middle point between two nanoslits, which is quite different from that seen in Fig.\,\ref{f3}.
In the lower-left panel of the figure, we consider an asymmetric configuration for double nanoslits
with $\delta=0$ and $\delta=0.5$ for the lower and upper nanoslits, respectively.
Although the hot spot still exists at the front surface for the reflection because of established electromagnetic coupling
between two different nanoslits, the bent-down candle-flame-like spot
at the upper-nanoslit exit becomes brighter than that at the lower-nanoslit exit for this asymmetric case. Furthermore, when these two asymmetric nanoslits are
moved apart from each other, as shown in the lower-right panel of the figure, the electromagnetic coupling between nanoslits no longer exists, which leads to a disappearance of the hot spot in the reflection
and a recovery of two deformed candle-flame-like spots. The lack of coupling between nanoslits also leads to the enhanced surface-plasmon propagation along the back surface of the silver film resulted from the
resonance between the incident light and the surface plasmon
at the metal-glass interface\,\cite{ref6}.
\medskip

The results in Figs.\,\ref{f3} and \ref{f4} correspond to a simultaneous illumination of all nanoslits by a $p$-polarized light in the plane-wave form.
On the other hand, we find from Fig.\,\ref{f5} that the nonlocal dynamic effects from embedded atoms in an unilluminated nanoslit, as a feedback, can affect the transmission through a separated illuminated nanoslit without embedded atoms.
Here, the bottom one of the two lower non-embedded nanoslits is illuminated by a narrow Gaussian light beam while the top unilluminated nanoslit is embedded with FIT atoms.
At $\lambda_0=620$\,nm for the two upper panels of this figure, the bottom illuminated nanoslit stays in an intermediate state\,\cite{ref1} having a medium direct transmission of light and a stiff local reflection,
while the top unilluminated and filled nanoslit
still acquires a relatively high cross-transmission coefficient much larger than that of the middle unilluminated nanoslit.
Moreover, the stiff reflection enhances the propagation of a resonant surface-plasmon mode at the metal-air (front) interface at this wavelength.
From the upper left panel of the figure, it is important to mention
that the electron dynamics in the top nanoslit has been mapped onto the dynamics for the direct transmission of light through the bottom non-embedded nanoslit,
where two Fabry-P\'erot-type peaks right above two dips for the top nanoslit are observed with their replicas in the bottom nanoslit due to inter-slit
coupling.  This coupling is facilitated by the resonant surface-plasmon mode localized at the metal-air interface.
This self-induced transmission for the embedded nanoslit is found to be
related to the sole-wave interference\,\cite{ref1} between the entry and exit edges of a nanoslit with a finite value for the average reflection coefficient (a large finesse increasing with the film thickness $2d$),
where the requirement for the Fabry-P\'erot constructive interference is fulfilled by a sharp increase in ${\rm Re}[\alpha_{\rm L}(\omega)]$ right above $\delta=0$ or $\delta=1$.
After $\lambda_0$ is switched to a longer wavelength at $990$\,nm, as shown in the two lower panels of this figure, the bottom illuminated nanoslit changes from an intermediate state to a passing state
with a large direct transmission of light and a softened local reflection.
In this case, the propagation of the previous resonant surface-plasmon mode at the metal-air (front) interface is suppressed at this longer wavelength,
while a new resonant surface-plasmon mode at the metal-glass (back) interface is induced.
As a result, two Fabry-P\'erot-type peaks for the top nanoslit become invisible owing to the cutoff of surface-plasmon propagation at the front interface and their replicas in the bottom nanoslit are also greatly weakened.

\subsection{Curvature effect in static control}
\label{sec3-2}

\begin{figure}[p]
\centering
\includegraphics[width=0.7\textwidth]{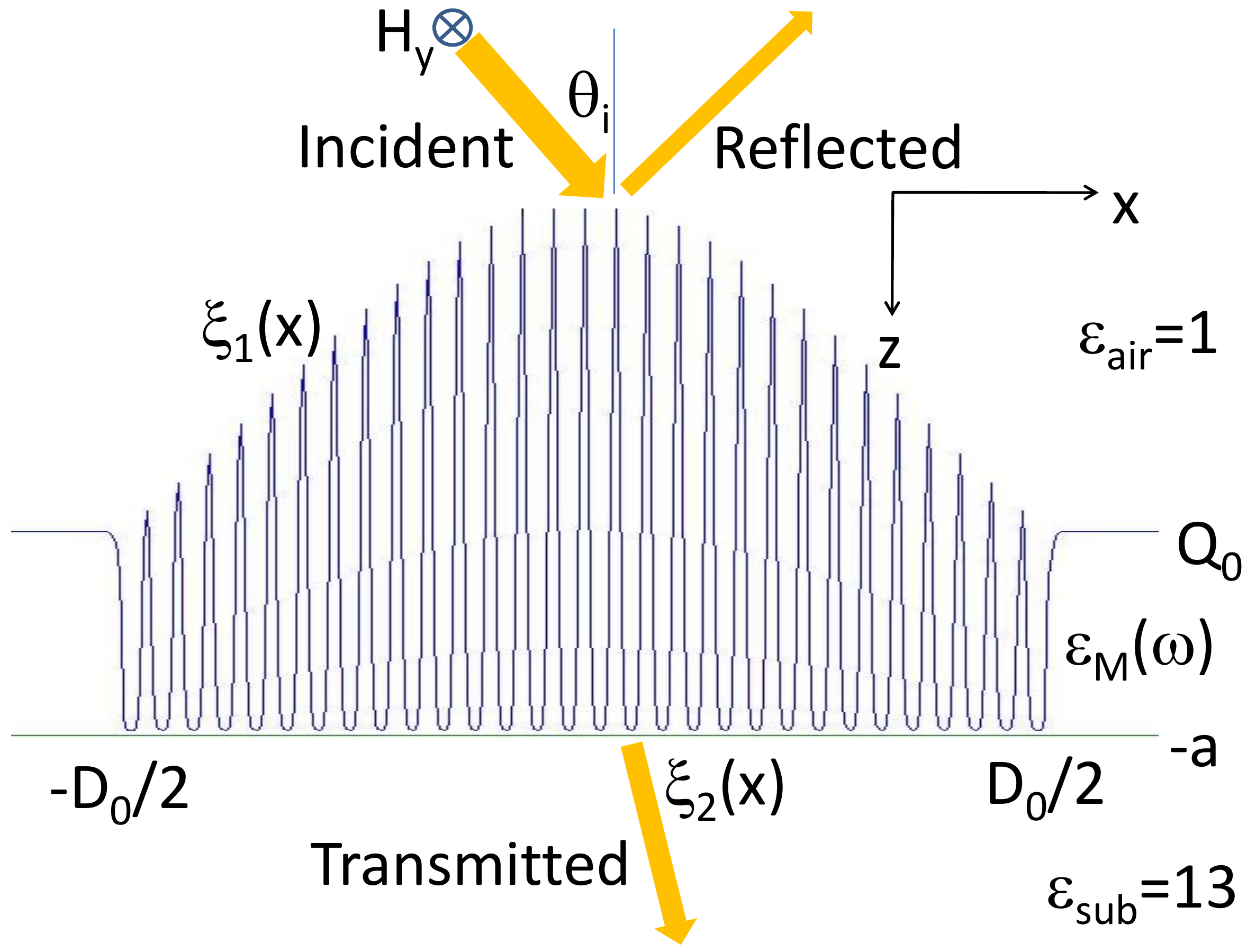}
\caption{\label{f6}
(Color online) Diagram representing a shaped metallic lens. A
Gaussian-overlapped profile $z=\xi_1(x)$ with an embedded slit array
is chosen for the surface on the entry side. A  flat surface
$z=\xi_2(x)$ is assumed for the exit side. A $p$-polarized plane-wave
light with out-of-plane magnetic-field component ${\bf H}=(0,\,H_y,\,0)$
is incident with  angle of incidence $\theta_i$  where the dielectric
constant is $\epsilon_{\rm air}=1$. The substrate dielectric constant
is $\epsilon_{\rm sub}=13$ on the exit side. A complex dynamical
dielectric function $\epsilon_{\rm M}(\omega)$ is employed for metal with optical loss included, where $\hbar\omega$ represents the incident photon energy. In addition, the metal-film thickness is $Q_0+a\approx Q_0$ ($a\ll\lambda_0$ with $\lambda_0$ being the incident wavelength), and $D_0$ represents the lens aperture size.
}
\end{figure}

\begin{figure}[p]
\centering
\includegraphics[width=0.9\textwidth]{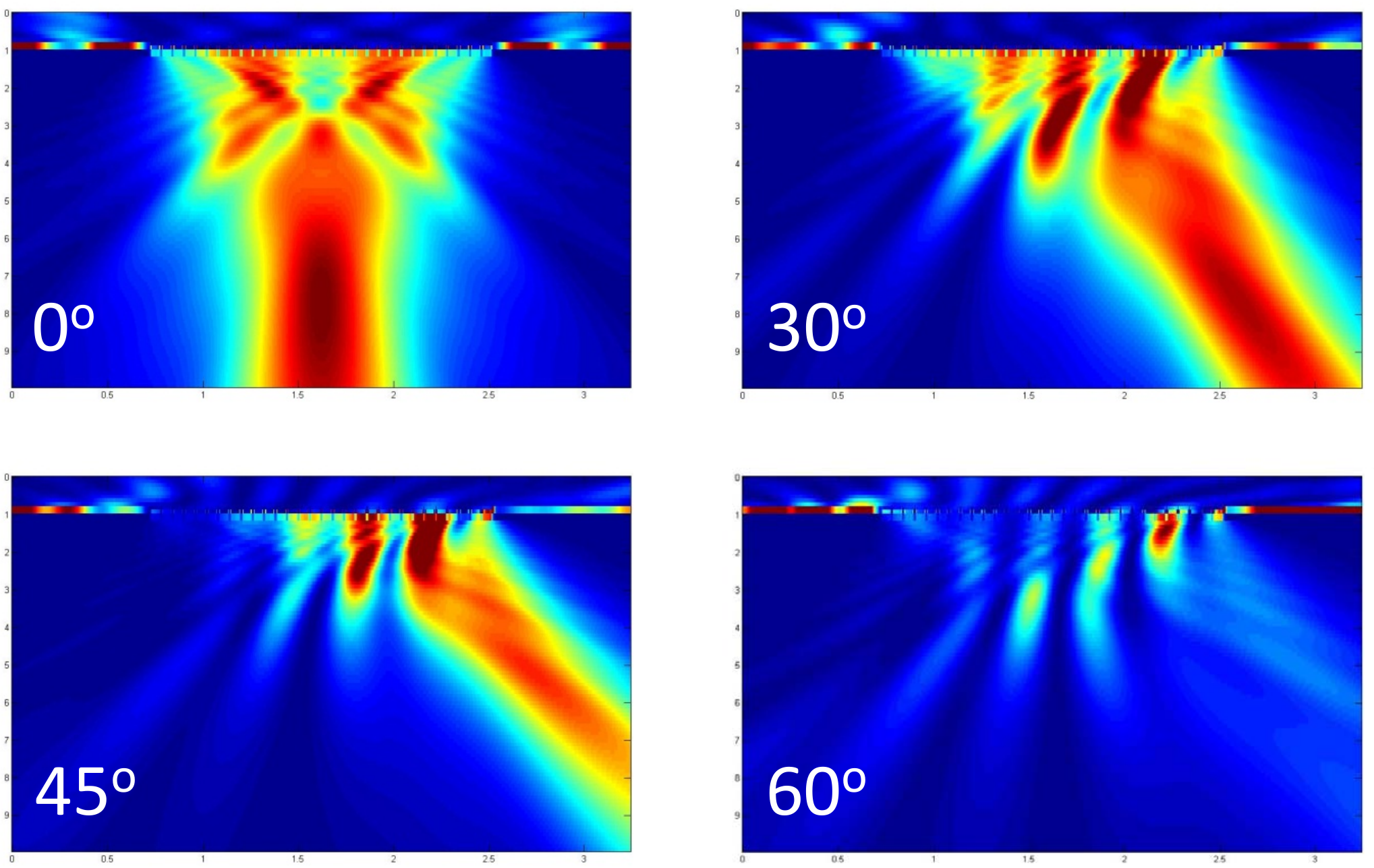}
\caption{\label{f7}
(Color online) Color plots of $|H_y(x,\,z)|^2$ for transmitted
light  through a flat and slit-embedded metallic film
 when $\lambda_0=540$\,nm. Various  angles $\theta_i$ of incidence
 are chosen. The color scales are from $0$ (blue) to $4.5$ (red)
 for $\theta_i=0^{\rm o},\,30^{\rm o}$, $0$ to $5$ for
 $\theta_i=45^{\rm o}$ and $0$ to $4$ for $\theta_i=60^{\rm o}$.
}
\end{figure}

\begin{figure}[p]
\centering
\includegraphics[width=0.9\textwidth]{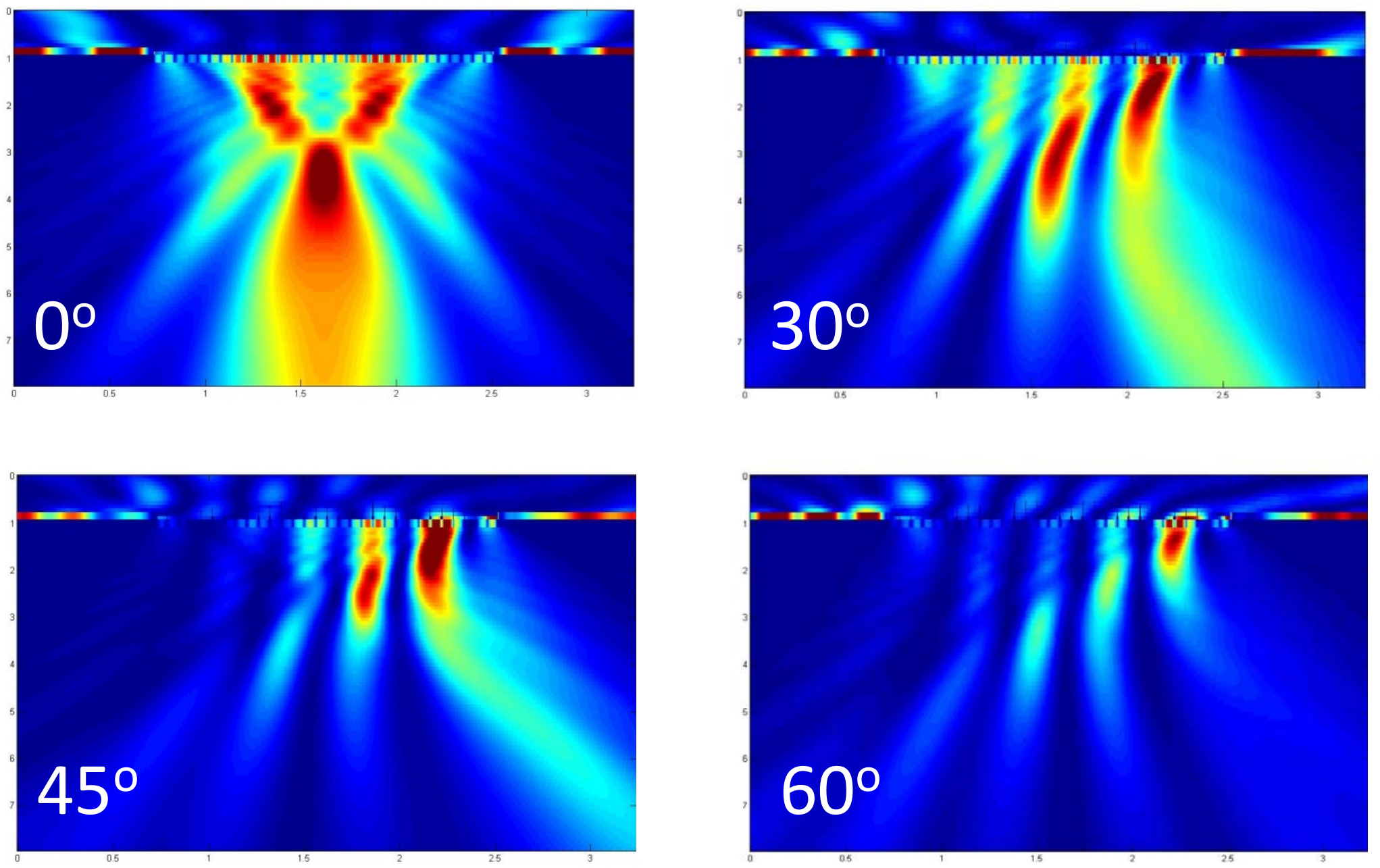}
\caption{\label{f8}
(Color online) Density plots of $|H_y(x,\,z)|^2$ for transmitted
light through a shaped and slit-embedded metallic film when
 $\lambda_0=540$\,nm for various angles  of incidence $\theta_i$.
The color scale ranges from $0$ to $5$ for
$\theta_i=0^{\rm o},\,60^{\rm o}$, $0$ to $6$ for $\theta_i=30^{\rm o}$
and $0$ to $7$ for for $\theta_i=45^{\rm o}$.}
\end{figure}

\begin{figure}[p]
\centering
\includegraphics[width=0.9\textwidth]{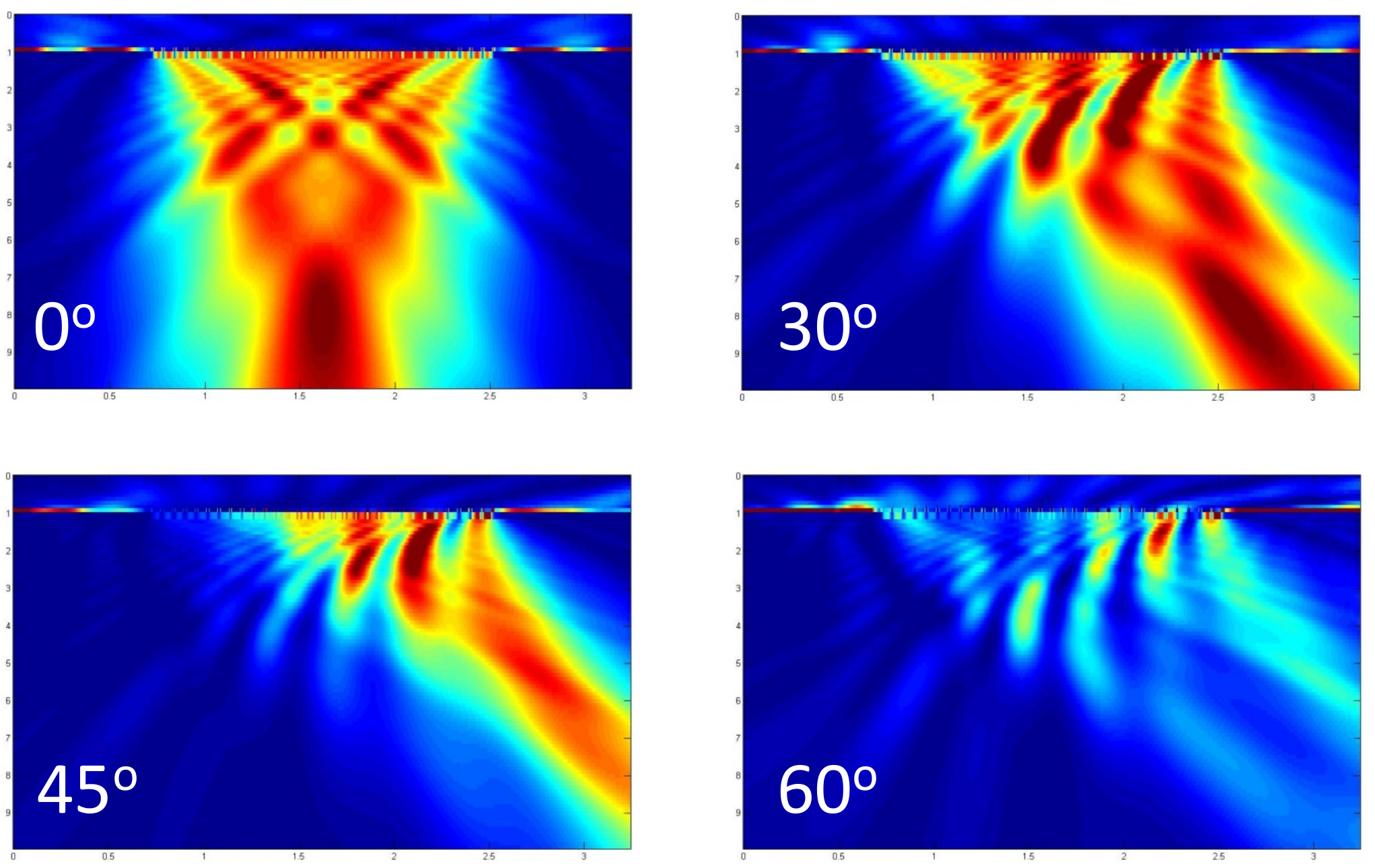}
\caption{\label{f9}
(Color online) Density plots of $|H_y(x,\,z)|^2$ for
light transmitted through a shaped and slit-embedded
metallic film. We chose $\lambda_0=540$\,nm with the half
values  for ${\cal A}_0$ and $Q_0$ used in Fig.\,\ref{f7}
and various angles  of incidence $\theta_i$. The color scale
ranges  from $0$ to $4$ for $\theta_i=0^{\rm o},\,30^{\rm o}$, $0$ to $5$ for $\theta_i=45^{\rm o}$ and $0$ to $3$ for for $\theta_i=60^{\rm o}$.
}
\end{figure}

\begin{figure}[p]
\centering
\includegraphics[width=0.9\textwidth]{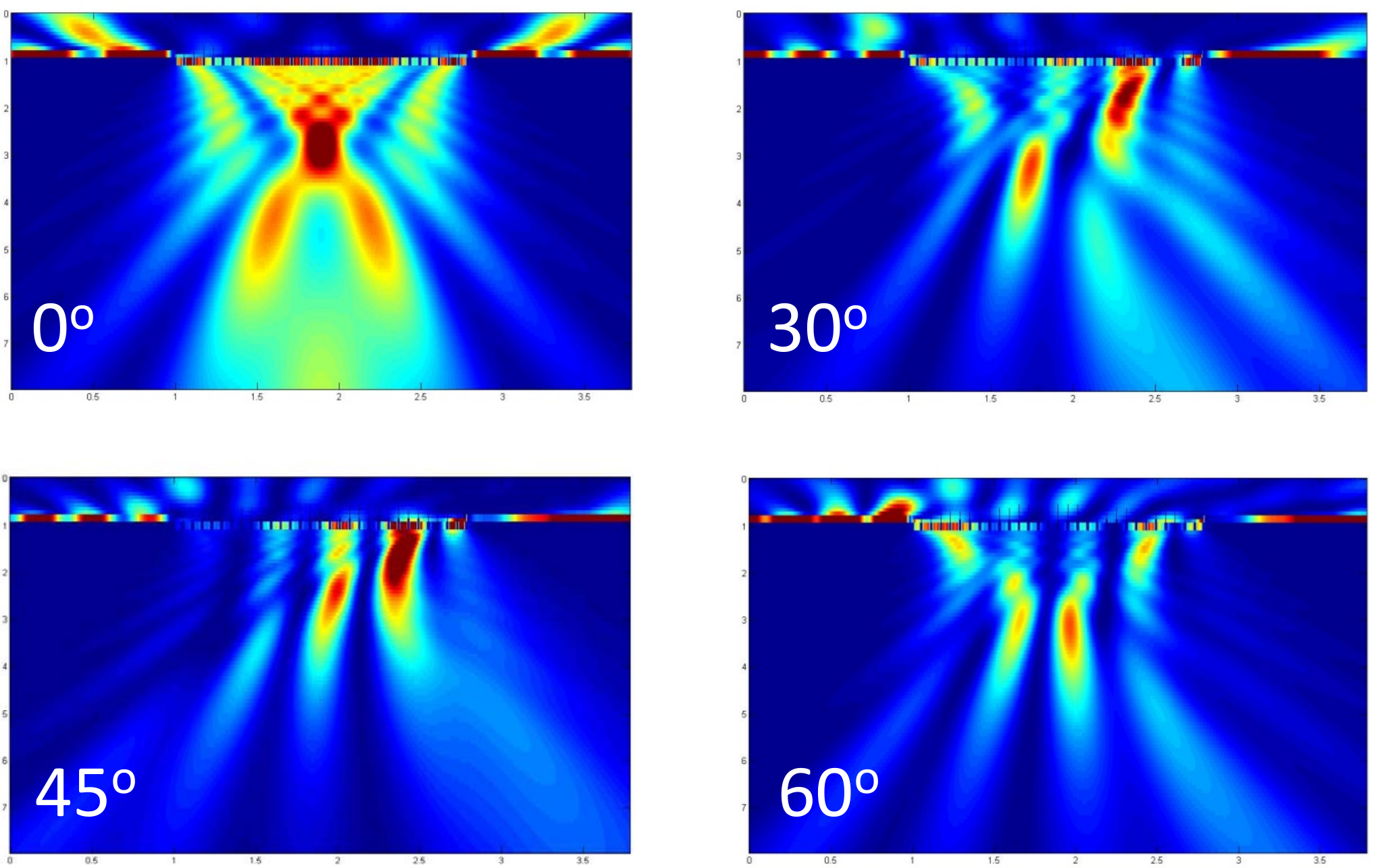}
\caption{\label{f10}
(Color online) Density plots of $|H_y(x,\,z)|^2$ for transmitted
light through a shaped and slit-embedded metallic film when
 $\lambda_0=630$\,nm for various angles of incidence $\theta_i$.
The color scale is  from $0$ (blue) to $5$ (red) for
$\theta_i=0^{\rm o},\,60^{\rm o}$, $0$ to $6$ for
$\theta_i=30^{\rm o}$ and $0$ to $7$ for for $\theta_i=45^{\rm o}$.
}
\end{figure}

For a shaped metallic microlens embedded with a finite nanoslit array, as shown in Fig.\,\ref{f6},
the parameters used for the numerical calculations in Figs.\,\ref{f7}$-$\ref{f10} are taken to be: $N=14$, $d=60$\,nm, $2b=6$\,nm, ${\cal A}_0=400$\,nm, $2\sigma_0=D_0=1.8\,\mu$m, $Q_0=152$\,nm, $a=4$\,nm,
$\epsilon_{\rm sub}=13$, and $\epsilon_{\rm air}=1$. Various  parameters
are given in the figure figure  captions.
\medskip

By neglecting the small surface-plasmon-polariton phase,
the spot-focusing condition to the leading-order
requires that the difference in the optical phases corresponding to any pair of nanoslit channels should be equal to a multiple of the incident wavelength $\lambda_0$\,\cite{ref7,ref8,ref9}.
Therefore, for a given arbitrary shape function $g(x)$ assigned to the front surface, it is found to satisfy the following constraint under normal incidence of light

\begin{equation}
g(x)=\frac{\sqrt{\epsilon_{\rm sub}}}{\sqrt{\epsilon_{\rm eff}(\lambda_0)}-\sqrt{\epsilon_{\rm air}}}\left(\sqrt{f_0^2+x^2}-f_0\right)-\frac{|m|\lambda_0}{\sqrt{\epsilon_{\rm eff}(\lambda_0)}-\sqrt{\epsilon_{\rm air}}}\ ,
\label{e5}
\end{equation}
where $x=jd$ for $j=0,\,\pm 1,\,\pm 2,\,\cdots,\,\pm N$, $m=0,\,\pm 1,\,\pm 2,\,\cdots$ is an integer, $f_0$ represents the designed short distance of a focused spot measured from the flat back surface,
$\sqrt{\epsilon_{\rm eff}(\lambda_0)}=\beta_{\rm slit}(\lambda_0)/k_0$
stands for the effective complex dielectric function of a nanoslit waveguide including the surface-plasmon contribution, $k_0=2\pi/\lambda_0$, and the complex wavenumber $\beta_{\rm slit}(\lambda_0)$ can be determined from
the nanoslit-waveguide boundary condition for a $p$-polarized plane wave.
By choosing $g(x)={\cal A}_0-\xi_1(x)\approx{\cal A}_0x^2/\sigma_0^2$ for $|x|\ll\sigma_0$, we find from Eq.\,(\ref{e5}) under $m=0$, to the leading order, a nanoslit-independent focal distance

\begin{equation}
f_0=\frac{x^2-{\cal F}_0^2(x)}{2{\cal F}_0(x)}\approx\frac{x^2}{2{\cal F}_0(x)}=\left[\frac{\sqrt{\epsilon_{\rm sub}}}{\sqrt{\epsilon_{\rm eff}(\lambda_0)}-\sqrt{\epsilon_{\rm air}}}\right]\frac{\sigma_0}{2{\cal K}_0}\ ,
\label{e6}
\end{equation}
where ${\cal F}_0(x)=[(\sqrt{\epsilon_{\rm eff}(\lambda_0)}-\sqrt{\epsilon_{\rm air}})/\sqrt{\epsilon_{\rm sub}}\,]({\cal K}_0x^2/\sigma_0)$, ${\cal K}_0={\cal A}_0/\sigma_0$, and a larger ${\cal K}_0$ value corresponds
to a small value for the focal distance $f_0$.
\medskip

Served as a purpose for comparison and a starting point for the static-control approach,
we display in Fig.\,\ref{f7} the calculated color maps of $|H_y(x,z)|^2$ for an unshaped gold film embedded with a finite nanoslit array at four different angles of incidence. In order to obtain a sharply-focusing spot
for the transmitted light at a very short distance $f_0$ in such a case,
the nanoslit-center positions $x_j$ with $j=\pm 1,\,\pm 2,\,\cdots$ (and $x_0=0$) for a $p$-polarized normal incidence must satisfy a relation, {\em i.e.\/},
$x_j^2=(|j|\lambda_0/\sqrt{\epsilon_{\rm eff}(\lambda_0)}\,)^2+2f_0|j|\lambda_0/\sqrt{\epsilon_{\rm eff}(\lambda_0)}$.
For a periodic nanoslit array with $x_j=jd$ and $j=0,\,\pm 1,\,\pm 2,\,\cdots,\,\pm N$, on the other hand,
we expect an elongated focusing spot in the $z$ direction, as seen in the upper-left panel of the figure at $\theta_i=0^{\rm o}$. Additionally, the elongated spot outside of the near-field region is
associated with the specular ($m=0$) mode, while those symmetric side robes with respect to $x_0=0$ are resulted from $m=\pm 1,\,\pm 2,\,\cdots$ higher-order diffraction modes within the near-field region.
When $\theta_i=30^{\rm o}$ in the upper-right panel of the figure, the specular mode is tilted accordingly similar to Snell's law. At the same time, the spatial
distribution of higher-order diffraction modes change and become asymmetric
with suppressed diffraction modes at the right side for $m=1,\,2,\,\cdots$. As $\theta_i$ is further increased to $45^{\rm o}$ in the lower-left panel of the figure, the specular mode is tilted even more, as expected,
and a series of hot spots in the near-field region is pulled back to the exit surface and squeezed towards the right edge of the nanoslit array at the same time.
Eventually, the specular mode becomes completely suppressed outside the near-field region in the lower-right panel of the figure at $\theta_i=60^{\rm o}$,
leading to a negative light-refraction pattern supported by three hot spots corresponding to one specular mode and two diffraction modes with $m=-1$ and $-2$ in the near-field region.
\medskip

For a shaped gold-film microlens embedded with a nanoslit array,
we present in Fig.\,\ref{f8} the color maps of $|H_y(x,z)|^2$ with various $\theta_i$ values. Compared with the results in Fig.\,\ref{f7}, the elongated focusing spot
originated from the specular mode in the upper-left panel of this figure is greatly shrunk in size and enhanced in strength simultaneously.
This is further accompanied by reduced intensities for diffraction modes with $m=\pm 1$ and $\pm 2$. Here, the enhancement in focusing capability
is attributed to the added compensation in the optical phase provided by the geometrical effect from a finite surface curvature for the shaped microlens.
The observed focal distance ($\approx 2.5\,\mu$m) at $\theta_i=0^{\rm o}$ is qualitatively consistent with the prediction ($f_0=2.8\,\mu$m) from Eq.\,(\ref{e6})
(by simply taking $\sqrt{\epsilon_{\rm eff}(\lambda_0)}\approx(\sqrt{\epsilon_{\rm air}}+\sqrt{\epsilon_{\rm sub}}\,)/2=2.3$ for the lowest $p$-polarized waveguide mode).
The deviation from an ideal spot shape in the upper-left panel of the figure comes from the nanoslit contribution in the outer region of $|x|\geq\sigma_0=0.9\,\mu$m.
The focusing quality is expected to be improved if only the central part of the nanoslit array is under an illumination by a narrow Gaussian light beam instead of a wide plane wave illuminating the entire array.
The hot specular mode outside the near-field region is greatly cooled down when $\theta_i\geq 30^{\rm o}$,
as can be verified by the comparison of the rest of the panels of this figure with their corresponding ones in Fig.\,\ref{f7}.
Moreover, the negative light-refraction pattern in the near-field region at $\theta_i=60^{\rm o}$, as shown in the lower-right panel of the figure, becomes much cleaner in vision after a weakly focusing spot from
the $m=-3$ diffraction mode enters into the game.
\medskip

As predicted by Eq.\,(\ref{e6}), the focal distance will depend on the surface curvature of a shaped microlens for the normal incidence of a $p$-polarized light.
We display the color maps of $|H_y(x,z)|^2$ in Fig.\,\ref{f8} for a reduced surface curvature (proportional to ${\cal K}_0$) as well as for a thinner gold film under different angles of incidence.
First, the crossover from a flat film to a shaped microlens can be understood by a comparison of the results in this figure with those in Figs.\,\ref{f7} and \ref{f8}, including the evolution of an
elongated focusing spot from the specular mode outside the near-field region in the $z$ direction,
the rapid development of side robes from the high-order diffraction modes in the near-field region, as well as the change in the focal distance.
Secondly, it is interesting to note the varied pattern of the near-field distribution from the upper-left-panel of the figure
and a new tiny focusing spot as well at a much shorter distance from the exit surface due to leaked higher-order waveguide modes in this case.
Finally, the reduced gold-film thickness makes it possible for the contribution to light transmission from the leakage of lossy higher-order waveguide modes,
which also show up in the coexistence of both the positive and negative light-refraction patterns in the near-field region at $\theta_i=60^{\rm o}$
in the lower-right panel of the figure.
Compared with Fig.\,\ref{f7}, we conclude that the focusing quality can be further improved if lossy higher-order waveguide modes can be fully filtered out by employing narrower nanoslits.
\medskip

In addition to the surface curvature, Equation\ (\ref{f6}) also indicates that the focal distance will depend on the incident wavelength $\lambda_0$ through the $\sqrt{\epsilon_{\rm eff}(\lambda_0)}$ factor.
Figure\ \ref{f10} shows the color maps of $|H_y(x,z)|^2$ at a resonant surface-plasmon wavelength\,\cite{ref6} (air-metal interface) at $\lambda_0=630$\,nm with four values of $\theta_i$.
For $\theta_i=0^{\rm o}$ in the upper-left panel of this figure, we find a reduced focal distance $f_0$, which is in agreement with the prediction $f_0\sim\sqrt{\epsilon_{\rm sub}}/\lambda_0$
from our previous study for a flat film embedded with a nanoslit array\,\cite{ref10}.
The strong coupling of the resonant surface-plasmon modes at two parts of the metal-air interface outside the microlens aperature
to the nanoslit waveguide modes produces intensive reflections at two edges of the array in opposite directions.
Moreover, the pattern of the near-field distribution for this case is significantly modified by a stronger, smaller and closer focusing spot resulting from the concentrated field at the central part of the array.
Here, the change in the near-field distribution pattern is closely related to the interference between two resonantly excited and counter-propagating surface-plasmon waves under the normal incidence.
However, the strong array-edge reflections are greatly weakened owing to the decreased interference effect from two counter-propagating surface-plasmon waves as $\theta_i$ increases,
as can be seen from the other three panels of the figure.
At $\theta_i=45^{\rm o}$ in the lower-left panel of the figure, a negative light-refraction pattern is observed, which is supported by both the specular mode and two diffraction modes with $m=-2$ and $-3$.
Interestingly, after removing the concentrated field at the middle part of the array at $\theta_i=60^{\rm o}$ in the lower-right panel of the figure,
a similar field-distribution pattern to that under normal incidence reoccurs because of strong reflection of one surface-plasmon wave at the left edge of the array,
where the diffraction modes with $m=-2$ and $-3$ at $\theta_i=60^{\rm o}$ play the role of those with $m=\pm 1$ at $\theta_i=0^{\rm o}$.

\section{Conclusions}
\label{sec4}

In brief, both dynamic and static approaches have been demonstrated for controlling the optical phase of a $p$-polarized light guided through a metallic film with a patterned surface in a deep subwavelength regime. For the dynamic control, the modulation of focused transmitted-light intensity has been observed within an extremely narrow frequency window. The surface-plasmon based electromagnetic coupling between the two nanoslits embedded with FIT atoms has shown additional control over the near-field distribution for symmetric and asymmetric configurations as well as for small and large separations between the two nanoslits. Moreover, an energy-transferring-process based light cross-transmission has been shown for the static control. In addition, the curvature-dependent effect on the position of a sharply-focused spot from a transmitted light passing through a Gaussian-shaped metallic microlens has been demonstrated at all angles of incidence.
\medskip

Physically, a new mechanism for near-field focusing of light has been predicted with its sharpness dynamically controlled by the frequency of the light in a very-narrow window. Technically, a mitigation scheme based on Green's second integral identity has been provided for avoiding drawback in numerically calculating near-field distributions by the finite-difference-time-domain method.

\begin{acknowledgments}
This research was supported by the Air Force Office of Scientific Research (AFOSR). ME and HM also would like to thank the supports from the AFRL Phillips and Space Scholars Programs, respectively.
\end{acknowledgments}

\newpage

\end{document}